\documentclass[preprint,12pt,authoryear]{elsarticle}

\usepackage{amsmath,amssymb}

\journal{Nano Communication Networks}

\begin{document}

\begin{frontmatter}



\title{Unimolecular FRET Sensors: Simple Linker Designs and Properties}


\author[ucd]{Shourjya Sanyal}
\ead{shourjya.sanyal@ucdconnect.ie}

\author[bu]{David F. Coker}
\ead{coker@bu.edu}

\author[ucd]{Donal MacKernan\corref{cor1}}
\ead{Donal.MacKernan@ucd.ie}

\cortext[cor1]{Corresponding author}

\address[ucd]{School of  Physics, University College Dublin, Belfield, Dublin 4, Ireland}
\address[bu] {Department of Chemistry\\ Boston University\\ 590 Commonwealth Avenue\\Boston, MA 02215, USA}

\begin{abstract}
Protein activation and deactivation is central to a variety of biological mechanisms, including cellular signaling and transport. 
Unimolecular fluorescent resonance energy transfer  (FRET) probes are a class of fusion protein sensors that allow biologists to visualize using 
an optical microscope whether specific proteins are activated due to the presence nearby of small drug-like signaling molecules, ligands or 
analytes.  Often such probes  comprise a donor fluorescent protein  attached to a ligand binding domain, a sensor or reporter 
domain attached to the acceptor  fluorescent protein, with these ligand binding and sensor domains connected by a protein linker. Various choices
of linker type are possible ranging from highly flexible proteins to hinge-like proteins. It is also possible to select donor and acceptor pairs 
according to their  corresponding  F\"oster radius, or even to mutate binding and sensor domains so as to change their binding energy in the 
activated or inactivated states.  The focus of the present work is the exploration through simulation of  the impact of such choices on sensor 
performance.
\end{abstract}


\end{frontmatter}





\section{Introduction}
\label{SEC:Introduction}
Measurement  of  biomarkers and ligands  are increasingly used to study transport, signaling and communication  in cells, and as 
diagnostics/prognostics of disease, or the presence of pathogens, allergens and pollutants in foods,  and the environment.   Accurate measurement 
in assays or cellular environments is important, and  protein based biosensors can be used in this context. But due to the molecular complexity of 
such sensors,understanding the features that determine their performance is  difficult both from the perspective of experiment, and 
detailed molecular dynamics. In the latter case this is due to the size of
the system to  be simulated and the associated time and 
spatial scales. To investigate such systems, at a qualitative level 
we use simple coarse grained models of proteins, and for critically important 
features requiring high accuracy, we employ advanced molecular dynamics, in particular rare-event methods.

Fluorescence (or F\"oster) resonance energy transfer 
(FRET) occurring between donor and acceptor fluorescent protein (FP) pairs can provide detailed spatio-temporal information about a wide range of 
biological processes.  
Typically, the FRET efficiency, $\cal I$ the average fraction of energy
transfer events per donor excitation event - falls off quickly with distance between the FPs near
the so called F\"oster radius, $R_0 \sim 5-7$ nm, thus offering a highly sensitive indicator of
spatial and temporal change between the FP pair. Biosensors incorporating FP pairs can be designed to respond to variations in local 
concentrations 
of target analytes (small signaling molecules or biomarkers), that change the internal structure of the biosensor,
bringing the FPs closer on average, which in turn can be observed optically through changes in the
FRET efficiency. 

Many unimolecular FRET based probes designed to monitor or report the local concentrations of
analytes, comprise a donor FP attached to a ligand binding domain, a sensor or reporter domain
attached to the acceptor FP, with these ligand binding and sensor domains connected by a linker (see Fig.1 for three examples).
When the ligand binding domain is activated due to 
the proximity
of a ligand or analyte (the so called ON state), an attractive interaction is turned on between
the binding and sensor domains causing them to come together, bringing their donor and acceptor
FPs closer. In the absence of the ligand/analyte (the OFF state), the domains should remain
further apart. Such spatial changes can be measured by changes in the FRET efficiency between
the FPs.

How well one can discriminate between the background or basal efficiency ${\cal I}_0$,
and changes in the FRET efficiency due to changes in the analyte concentration close to the
sensor is determined by the signal-to-noise ratio  $({\cal I} - {\cal I}_0)/{\cal I}_0 = \Delta
{\cal I}/{\cal I}_0$, and is of critical importance in sensor design.  A related quantity 
is
${\cal J} = \frac{ \sigma({\cal I} - \cal{I}_0)} {\mu({\cal I}-{\cal I}_0)}$ (i.e. the fractional error in the gain $\mu({\cal I}-{\cal I}_0)$) 
which is simply related to the so called Z' factor used to 
characterise the quality of a sensor.  
In particular, one  can easily  show (making
reasonable assumptions) that the fractional error in the ligand/biomarker  concentration predicted from calibrated FRET measurements   is
proportional to $\cal J$.  Here $\mu$ and $\sigma$  denote the mean and the variance. This allows the effect of changes in the sensor design to be 
easily  related   to the  accuracy at which  concentrations of target ligands/biomarkers can be measured.

\begin{figure}[!t]
\centering
\includegraphics[width=6.5cm]{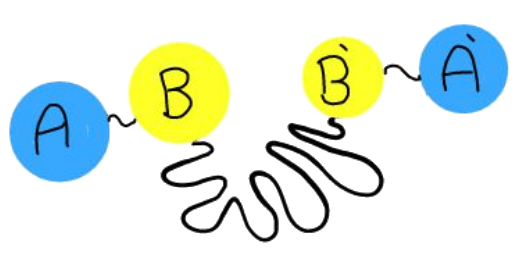} \hspace{0.5cm}
\includegraphics[width=6cm]{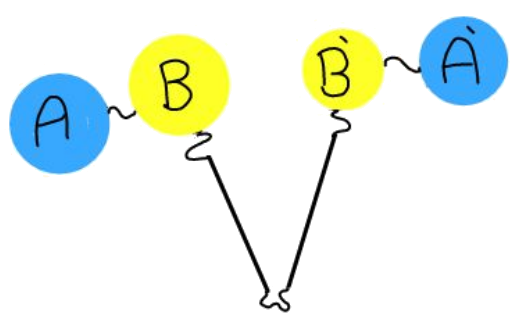} \\
\includegraphics[width=12cm]{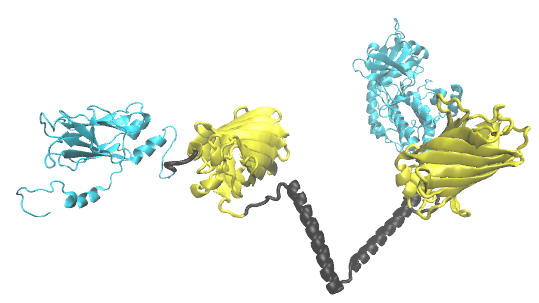}
\caption{Top-left is a schematic illustration of a unimolecular sensor  where a flexible linker is used to connect protein modules B and B'. 
Top-right corresponds to the case where the flexible linker is replaced by a free hinge type protein. The bottom figure is an example of 
a PKA sensor where the yellow cylindrical-like proteins flanking the hinge are FP's;  and the sensing units are the  PKA substrate (far left)   
and corresponding consensus protein respectively (far right). When the  PKA substrate  is phosphorylated by PKA, it will bind to the consensus 
protein. Note frequently the order of A,A' and B,B' is interchanged.}
\label{fig_sim}
\end{figure}

The choice of molecular linker used to connect the components B and B' of the biosensor depicted in  the top panels of Fig. \ref{fig_sim}
can have a strong influence on its overall performance\cite{lissandron2005}. In this current work we first model the flexible linker system 
developed by \cite{komatsu2011}  using a variable numbers of repeat units of the form (SAGG)$_n$ to design a FRET biosensor for Kinases 
and GTPases. We then compare these results with idealized models of hinge type linkers built using  $\alpha$-helical proteins. This will allow 
four general design questions to be considered. First, can a simple mechanistic model of the Komatsu sensor capture the salient features observed 
in experiment?  Second, for unimolecular sensors, is there an advantage in replacing the flexible linker peptide of \cite{komatsu2011} with a 
hinge peptide? Third, to enhance precision of measurement, is it in  principle better to increase of decrease the the F\"oster radius of 
fluorescent proteins? Fourth, is precision enhanced or reduced if the binding energy of the ligand and sensor domains is attractive or repulsive 
in 
the absence of the target ligand?

\section{Methods}
To analyse experimental FRET microscopy results, and more
generally, to explore idealized design motifs for chromophore - linker - chromophore systems,
we have built simplified models of unimolecular FRET probes, represented by two macro-particles
joined by an idealized linker. One spherical macro-particle represents the donor fluorophore
attached to the ligand binding
domain and the other represents the acceptor fluorophore attached to the signaling domain. The
macro-particles are connected to either end of a peptide linker, which may be flexible, or
``hinge-like'' modeling strong secondary structures such as a pair of flexibly connected or
hinged alpha helices \cite{boersma2015}.

The spherical macro-particles interact through a pair potential of the form
\begin{equation}
V(R) = V_s(R) +  V_{\ell}(R)
\label{eq:V}
\end{equation}
where the first term denotes binding between the macro-particles due to the presence of
the target ligand/analyte, and the second term is an interaction specific to each linker type.
When the ligand binding domain is in the OFF or basal state,  $V_s$ ensures that the spherical
macro-particles cannot overlap, $V_s(R) = \infty$ if $R < \sigma$ and is zero otherwise. When
the signal domain is in ON state, $V_s$ has, in addition to this excluded volume interaction,
an attractive square well interaction of depth
$\epsilon$ for $\sigma < R < \sigma + \Delta$, where $R = |\vec{R}_2 - \vec{R}_1|$ is the
distance between the macro-particles. The protein diameter defined as $\sigma$, is used as
the unit of  length. The F\"oster Radius $R_{0}$
is assumed to be 2.5 times greater than $\sigma$, and $\Delta$, the width of the attractive
well, is set at 0.2$\sigma$. The binding energy in the ON state is specified by $\epsilon$,
which is given in terms of $k_BT$, where $T$ corresponds to physiological temperature of 309K.

\subsection*{Flexible linker}\label{subsec:flex}
With the flexible linker model the two spherical macro-particles with excluded diameter $\sigma$
represent the FRET fluorophores and their associated proteins. The form of $V_s$ has been defined
above, and the linker part of the interaction is a simple isotropic pair potential with the form

\begin{equation}
V_\ell(|\vec{R}_2 - \vec{R}_1|)
= 
\begin{cases} 
\infty & \mbox{ if }  R > {\cal L} \\
   0   & \mbox{otherwise } 
\end{cases}
\end{equation}
\label{eq:simplelinker}
Geometrically this can be visualized as two non-overlapping macro-particles free to move inside a
sphere of diameter $\cal L$. To compare FRET efficiency predictions of the simple flexible linker
model with experiment where the linker length is given as the total number of residues $\cal N$,
it is necessary to relate $\cal N$ to $\cal L$. This was done through their corresponding mean
square center-to-center distances, $\langle R^2\rangle$ (see appendix \ref{SEC:A2} ). \cite{sanyal2016}

For the experimental system, if the linker is sufficiently flexible, the center-to-center distance
can be approximated as a Gaussian random walk for which
\begin{equation}
\langle R^2 \rangle = D_0^2 + C_{\infty} {\cal N} b_0^2,
 \end{equation}
where $D_0$ is the diameter of the macro-particles, $C_\infty $ is the characteristic
ratio, and $b_0$ is the
distance between consecutive C-$\alpha$ atoms in the peptide chain\cite{sanyal2016,Merkx2006}.
The corresponding Kuhn length of the model is given by $C_\infty b_0$, and is applicable
to flexible peptides. As we have seen in Fig. \ref{fig:flexible}, we find that once
$\cal L$ is related to $\cal N$ in this way, there is very close correspondence between the
prediction of the model and the experiment results with only slight differences occurring when
the linker is short arising as a consequence of departure from the ideal Gaussian chain behavior, 
see  Fig. S1(b) in \ref{SEC:A1} for further details. \cite{sanyal2016}

\subsection*{Spherical hinge linker}
For the spherical hinge linker, $V_{\ell}(|\vec{R}_2 - \vec{R}_1|)$ simply corresponds to two rods
of equal length
$\frac{\cal L}{2}$ connected by a freely rotating joint at  the origin, or equivalently to
the constraints
$
{x_1}^2 + {y_1}^2 + {z_1}^2 = \frac{L^2}{4} = {x_2}^2 + {y_2}^2 + {z_2}^2,
$
which geometrically can be viewed as two non-overlapping macro-particles free to move on
the surface
of a sphere of diameter $\cal L$.

\subsection*{Circular hinge linker}
For the circular hinge linker model,  $V_{\ell}(|\vec{R}_2 - \vec{R}_1|)$ is similar to that of
the spherical hinge, with the additional
constraints that $z_1 = 0 = z_2$, which geometrically corresponds to two non-overlapping
macro-particles free to move on a circle
of diameter $\cal L$.

\subsection*{Observables and sampling procedure}
The distance dependence of the FRET efficiency is approximated by the expression,
\begin{equation}
I(R) = {1 \over 1 + (R/R_{0})^6}
\label{eq:r0}
\end{equation}
with the F\"oster radius $R_{0} \sim 5-7$ nm giving the distance at which the energy transfer
efficiency is $50\%$ and
$R$ is the distance between the spherical macro-particles. To calculate the efficiency, as
measured in the experiment,
we compute its expectation value so ${\cal I} = \langle I(R) \rangle$ where the angle brackets indicate the corresponding average over the 
Boltzmann distribution either the the OFF and ON states. 
The F\"oster radius $R_0$  depends on various quantities including: the fluorescence quantum yield of the donor
in the absence of the acceptor, the refractive index of the medium, and the dipole orientation
factor $\langle \kappa^2 \rangle$ (see section \ref{SEC:A2}). We use the Monte Carlo simulation approach 
\cite{metropolis1953,Frenkel1996,corry2005} to 
estimate the statistical properties of each model. Further details of the observable, underlying theoretical assumptions
and the sampling procedure are given in \ref{SEC:A2}.

\section{Results}
\label{SEC:results}
The influence of different geometrical/structural properties of linkers on the FRET efficiency
is explored here using simple statistical mechanics models and Monte Carlo simulations.

\subsection*{Comparison of Simulation \& Experiment for the flexible linker}
\begin{figure}[!t]
\centering
\includegraphics[height =5cm]{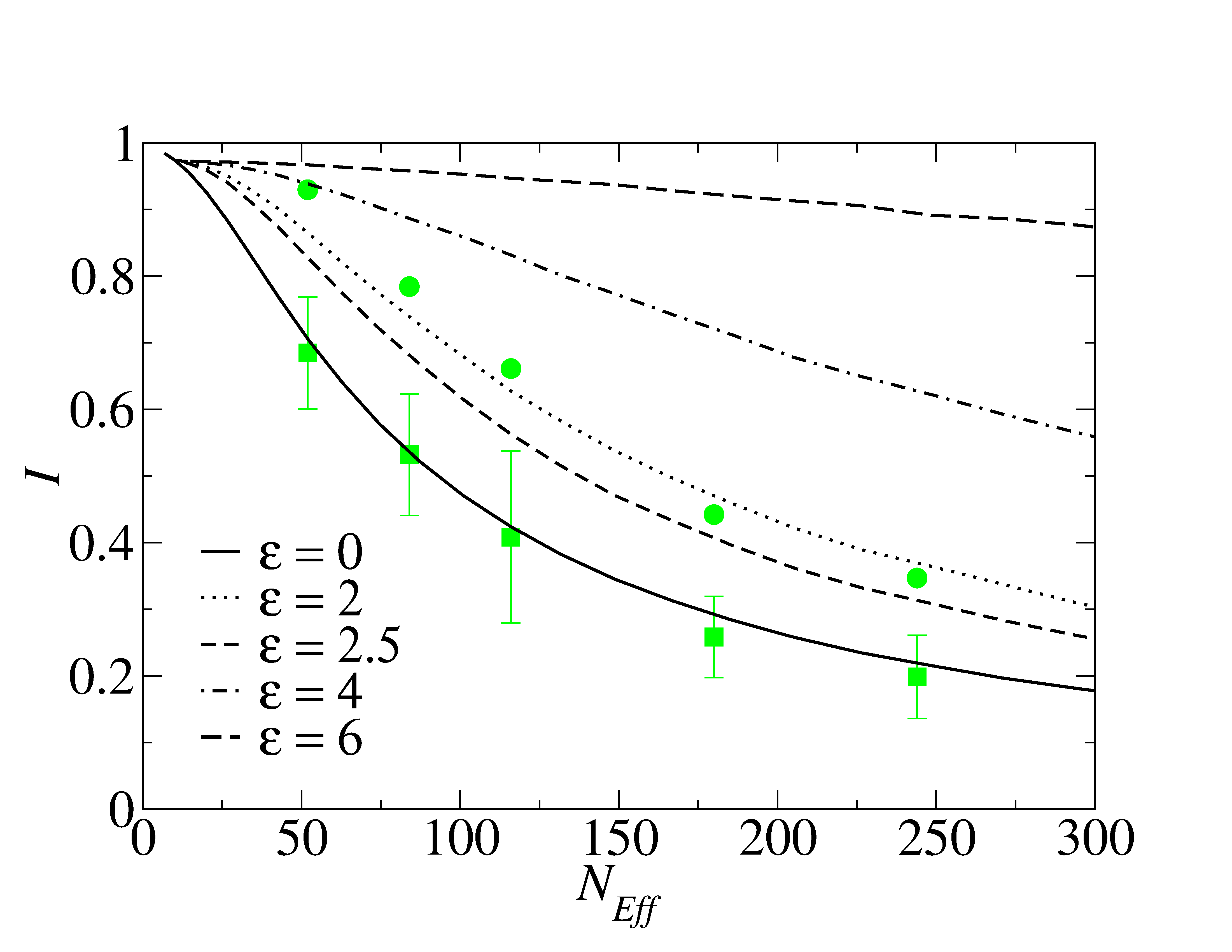} (a) \hspace{-.25cm}
\includegraphics[height =5cm]{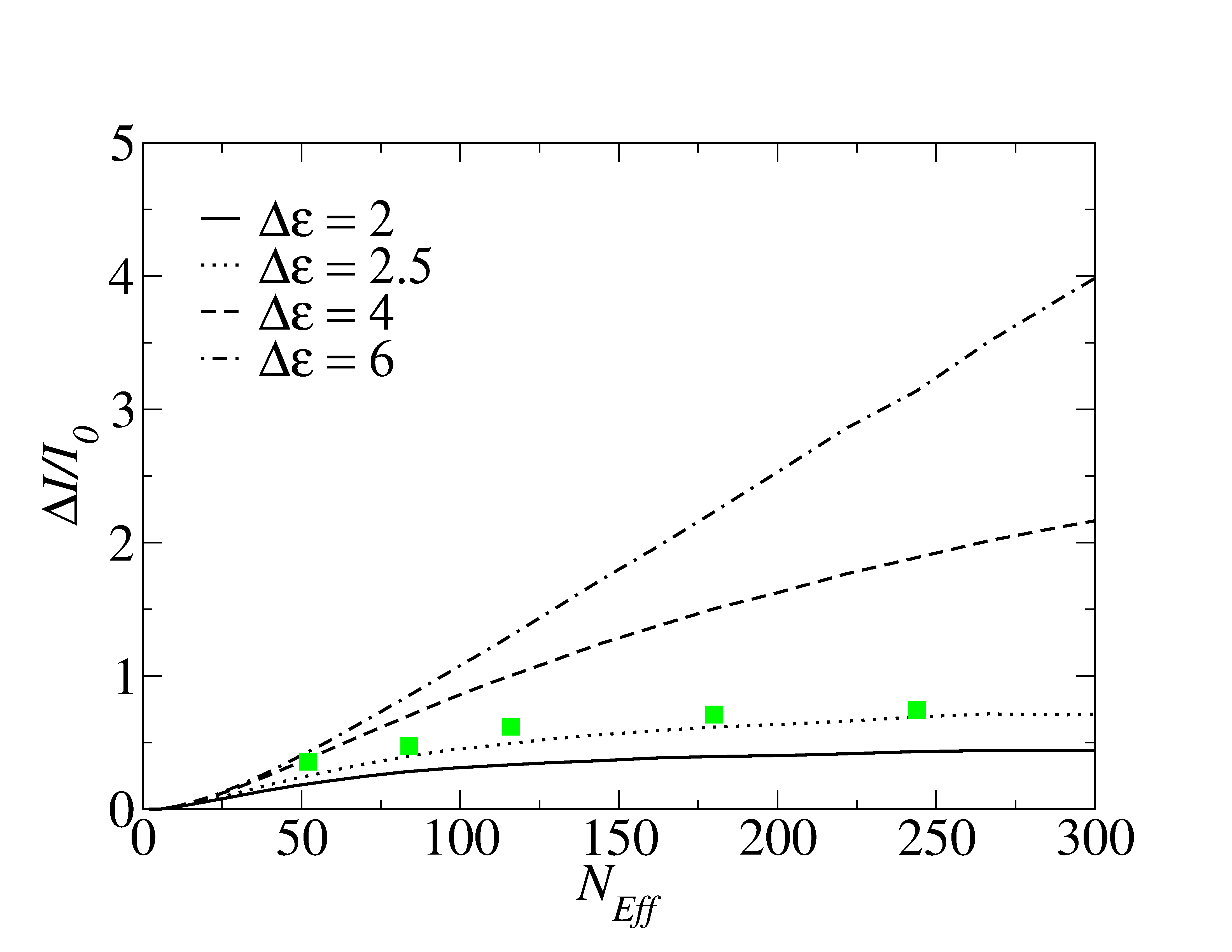}  (b) \\
\includegraphics[height =8cm]{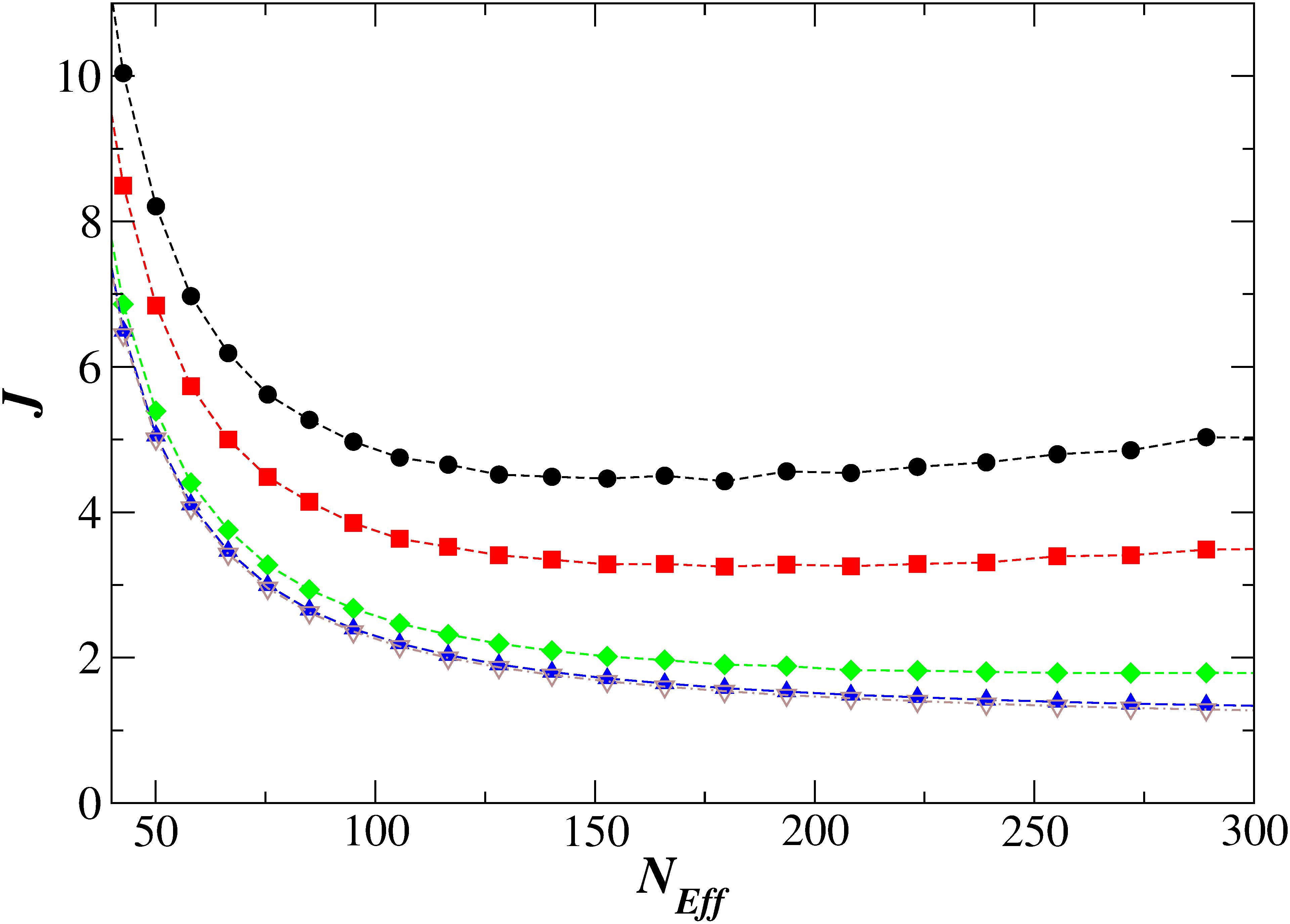} (c) 
\caption{(a) FRET intensity $\cal I$ of the flexible linker model as a
function of number of linker residues. Results are presented for five different values of the binding energy 
$\epsilon$, where $\epsilon = 0$ corresponds to the basal case.  Data from experiments \cite{komatsu2011}, in the OFF or basal state
(filled green squares) and in the ON state (filled green circles) are superimposed on the
theoretical predictions. 
(b) Corresponding signal-to-noise $\Delta{\cal I}/{\cal I}_0$ for the 
theoretical model overlaid with the experimental signal-to-noise ratio data (filled
green squares). (c) ${\cal J}$ of the flexible linker model as a function of effective number of residues 
$N_{Eff}$. The lower the value of $\cal J$, the more accurate the sensor, where each curve corresponds to a value of 
$\epsilon$, black bullet 2; red square 2.5;  blue lozenge 4; and green triangle 6  (in units of $k_BT$).  }
\label{fig:flexible}
\end{figure}

In Fig.\ref{fig:flexible} we compare the results of our flexible linker model simulations
with the experimental findings for both signal (a), and signal-to-noise ratio (b) obtained by 
\cite{komatsu2011} as a 
function of linker length. The comparison  give consistent estimates for the
ON state binding energy for this particular experimental system of $\epsilon=2.5$ $k_B T$. 
In panel (c) $\cal J$ is plotted  as a function of effective number of residues 
$N_{Eff}$.
\subsection*{Comparison of sensor performance for flexible and hinge linkers}
\begin{figure}[!t]
\centering
\includegraphics[height =4.5cm]{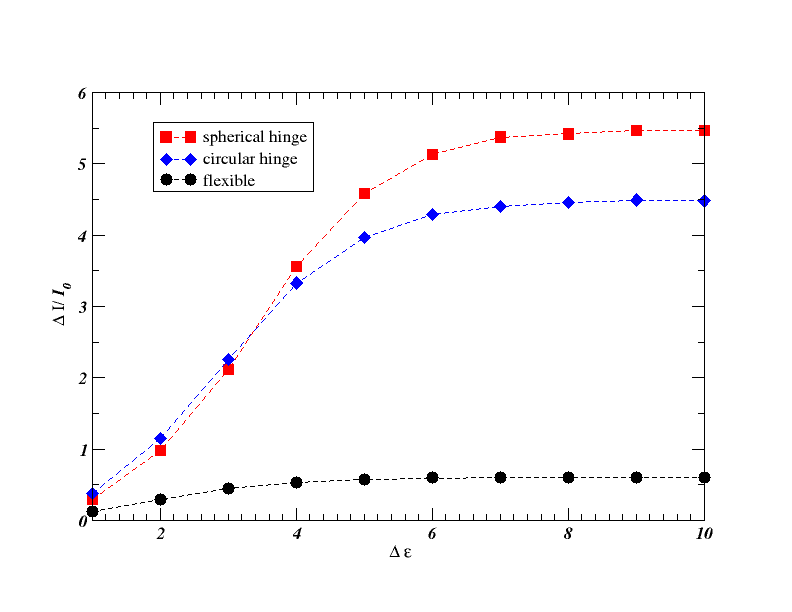} (a) 
\includegraphics[height =4.5cm]{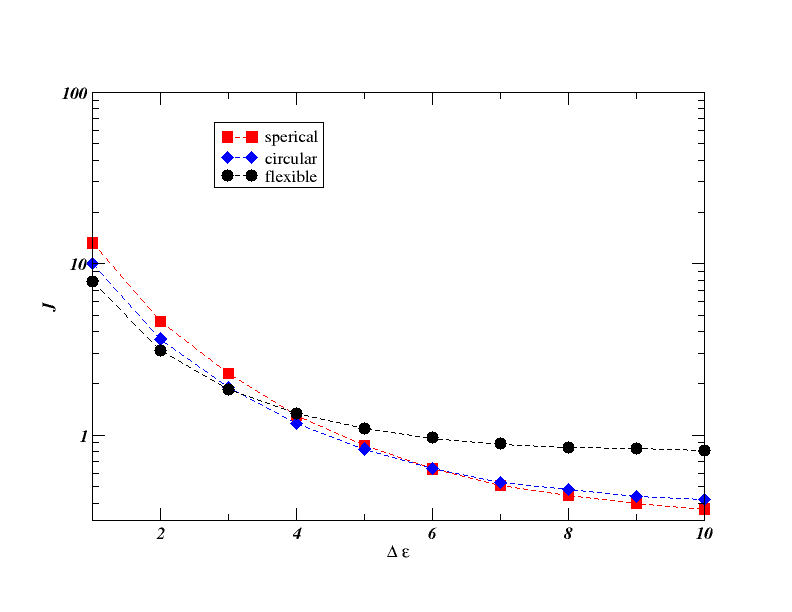} (b)
\caption{$\Delta{\cal 
I}/{\cal I}_0$ (left panel) and $\cal J$ (right panel) are plotted for the spherical hinge, circular hinge and flexible linker sensors 
respectively for $ {\cal L} = 3.48 \, \sigma $. For flexible linkers of \cite{komatsu2011} this correspond to 116 amino-acids/residues, and for 
hinge linker sensors, this correspond to  length for each arm of 4.2 nm or equivalently 28 amino-acids, each arm being an alpha helix.}
\label{fig:flexibleVhinge}
\end{figure}

To compare the performance of sensors when the flexible linker between B and B' is replaced by a free hinge, we demanded  that the arms of each 
hinge consist of about  28 residues (alpha helices of this length can be selected that are structurally stable) and that the flexible linker 
correspond to the optimal linker of \cite{komatsu2011} , which was 116 residues in length. Fig \ref{fig:flexibleVhinge} (a)  shows that under such 
assumptions,  the signal to noise ratio's where free 
hinge linkers are used  instead of flexible linkers are significantly higher.  Fig \ref{fig:flexibleVhinge} (b) which plots $\cal J$ indicates 
that hinge linker based sensors for moderate and high binding energies are likely go give rise to much more precise sensors. Examples of such 
hinge proteins include those reported  by \cite{boersma2015},and   behave as  free 
spherical hinges (the detailed free energy simulation results are not displayed here due to space limitations).  
\subsection*{Role of F\"oster radius on sensor performance}
\begin{figure}[!htb]
\centering
\begin{tabular}{c}
\includegraphics[height =8 cm]{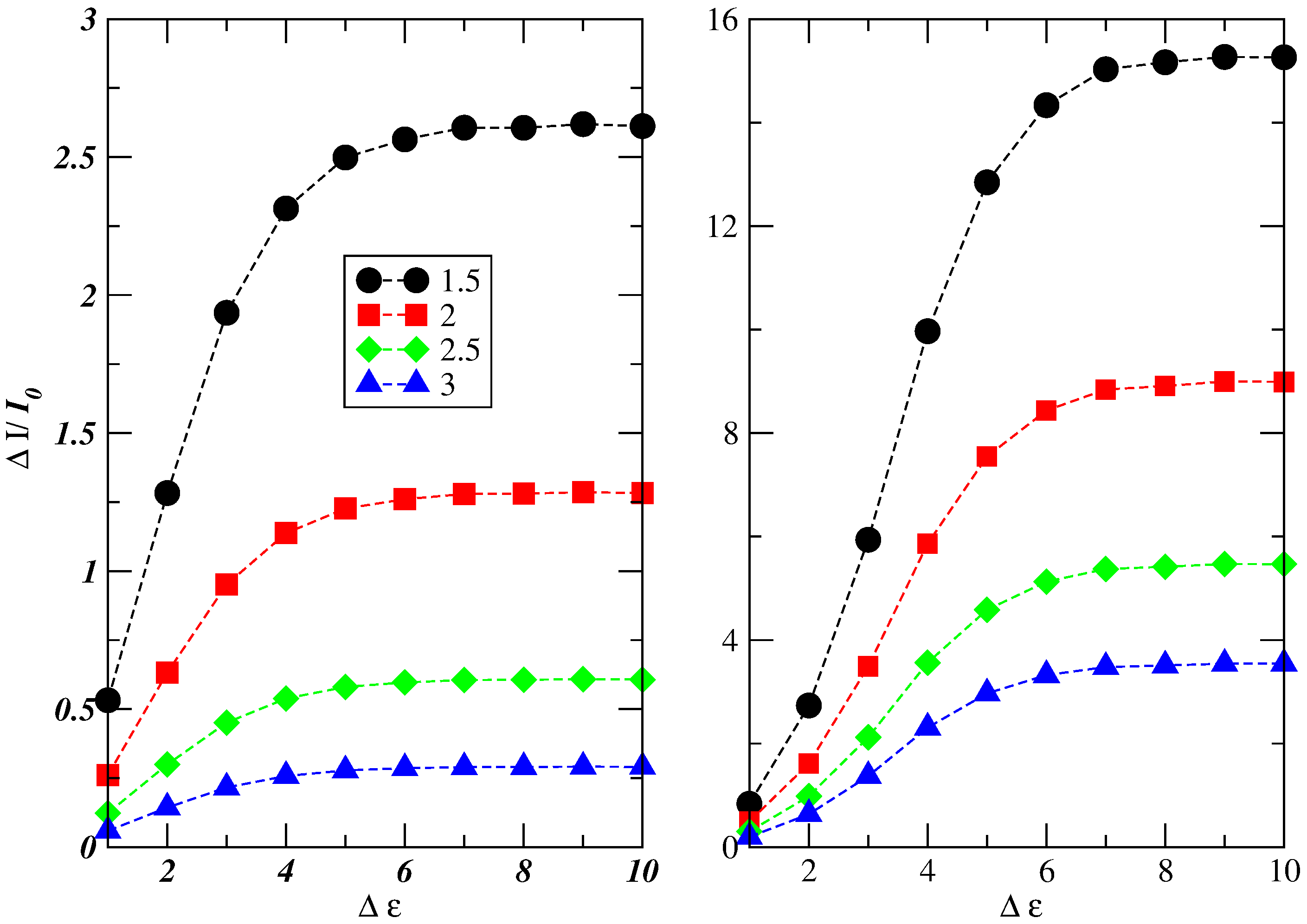} \\
\includegraphics[height =8 cm]{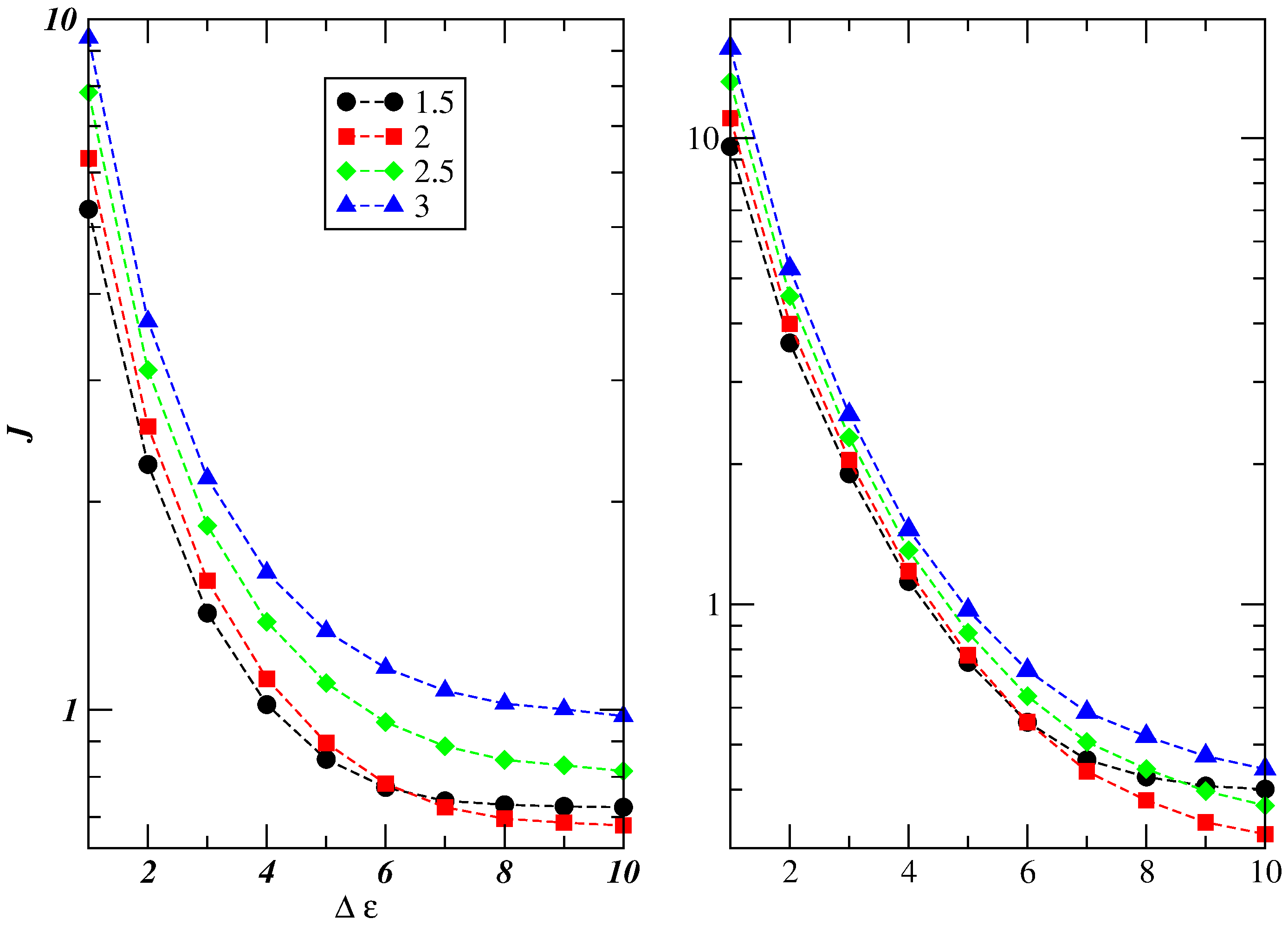}
\end{tabular}
\caption{Signal-to-noise $\Delta{\cal I}/{\cal I}_0$ (top panels) and fractional error in the gain $\cal J$ (bottom panels) as a function of  
F\"oster Radius $R_0$, with  ${\cal L} = 3.48 \, \sigma$ (corresponding to the flexible linker sensor consisting of 116 residues of  
\cite{komatsu2011} (left panels) and spherical hinge sensors (right panels) where each arm of the hinge is of length 4.2 nm).  }
\label{fig:compare}
\end{figure}

While the FRET efficiency for all systems must increase with increasing $R_{0}$, as is  observed
in experiments, \cite{hink2003} it is not clear how the signal-to-noise ratio should vary. Calculation results for our model system (see Fig. 
\ref{fig:compare}) show  that $\Delta{\cal I}/{\cal I}_0$ decreases and $\cal J$ increases dramatically with increasing $R_{0}$.   
This suggests that trying to increase signal to noise by increasing the $R_{0}$ 
can be counter-productive. Instead, reducing the F\"oster radius where possible is likely to significantly enhance sensor accuracy.
\subsection*{Effect of non-zero basal binding energy}
\begin{figure}[!htb]
\centering
\begin{tabular}{c}
\hspace{-2cm}
\includegraphics[height =9.0 cm]{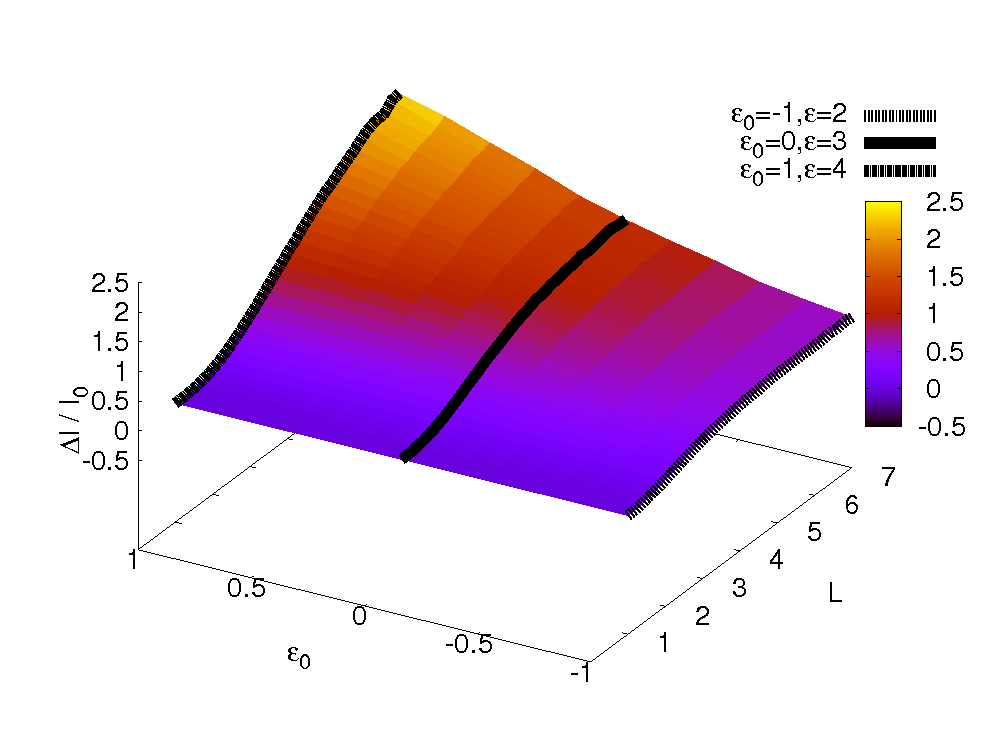} (a) \\ 
\hspace{-2cm}
\includegraphics[height =4 cm]{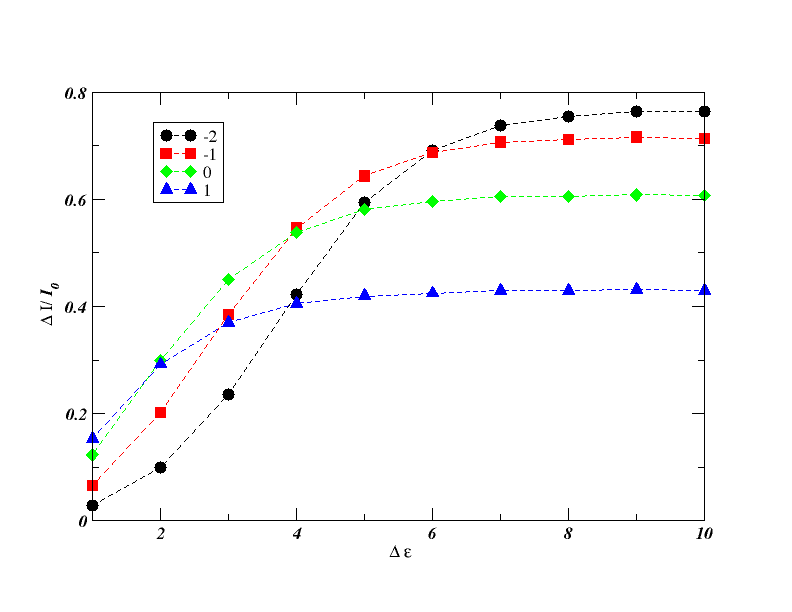}(b) 
\includegraphics[height =4 cm]{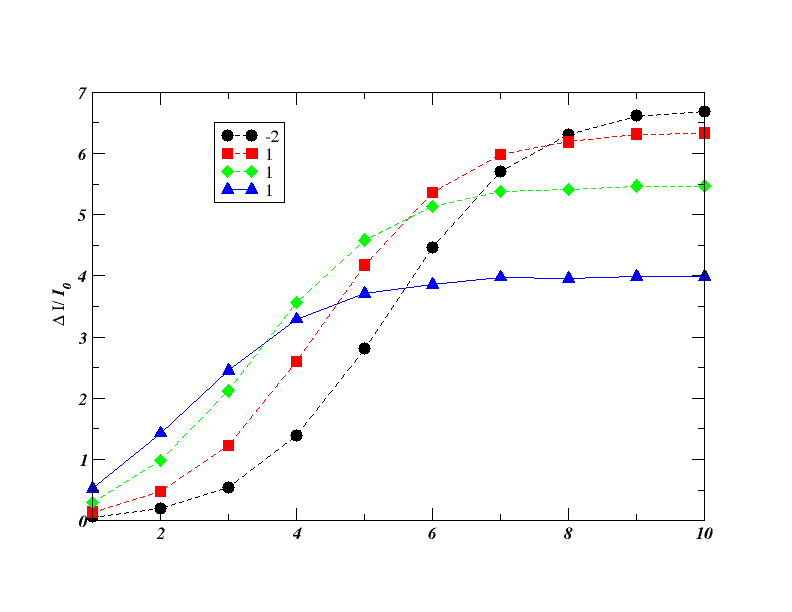}(c) \\ 
\hspace{-2cm}
\includegraphics[height =4 cm]{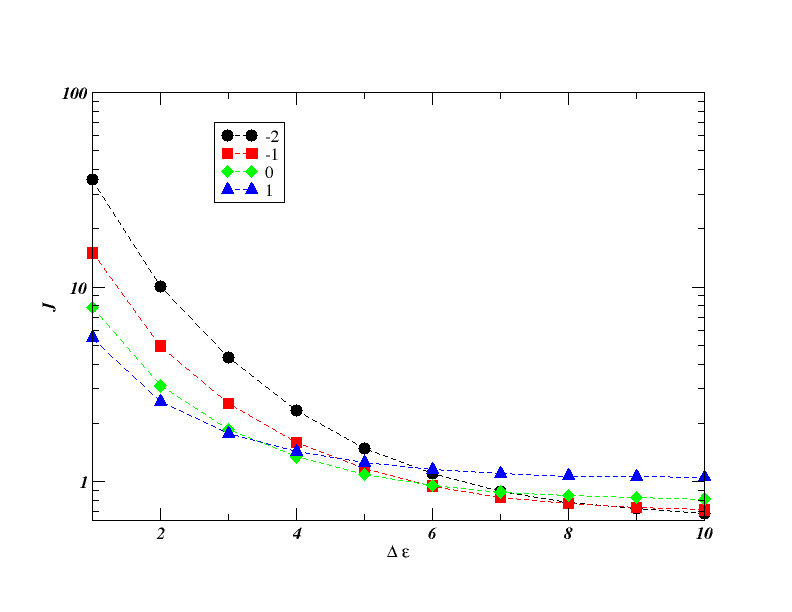} (d) 
\includegraphics[height =4 cm]{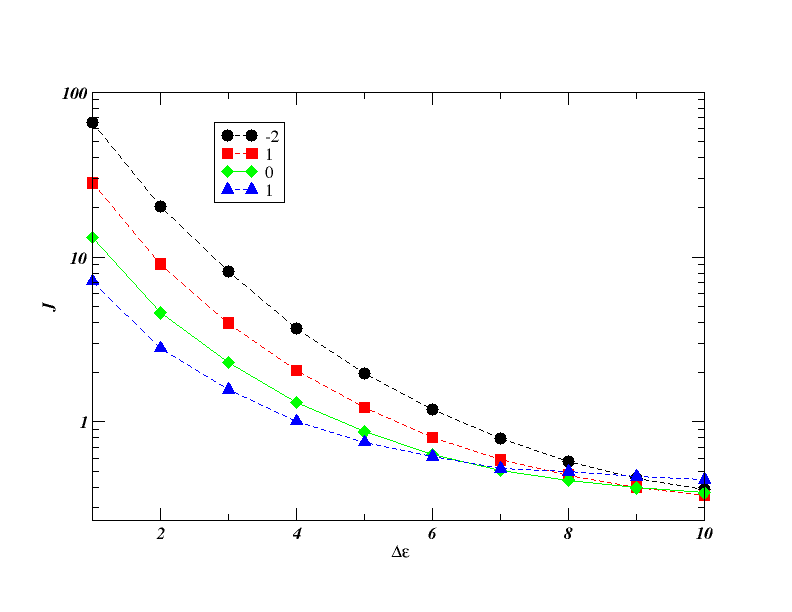} (e)
\end{tabular}
\caption{(a) $\Delta{\cal I}/{\cal I}_0$ and $\cal J$ of the flexible linker, spherical hinge linker and circular hinge linker  model sensors as a 
function of basal binding energy $\epsilon_0$ varying from -2k$_b$T to 1k$_b$T,  and  in the top panel(a)  for the flexible linker sensor model 
also as a function $\cal L$ in units of macro particle diameter $\sigma$. Dependence of the $\Delta{\cal I}/{\cal I}_0$ (figures a, b and c) and 
$\cal J$  (figures d and e) respectively on 
$\Delta \epsilon$  for ${\cal L} = 3.48 \,\sigma$  and representative values of  $\epsilon_0$ for the flexible linker, 
spherical hinge linker sensors respectively.}
\label{fig:basalfretvar}
\end{figure}

For simplicity, we have assumed that in the OFF or basal state the only interaction between the ligand binding and sensor
domains is a hardcore repulsion preventing their  overlap, and correspondingly set the binding energy $\epsilon$ between spherical
macro-particles to zero in the OFF state. However, an additional attractive or repulsive interaction is possible even in the absence 
of the ligand/analyte. This can be modeled as non-zero basal binding energy  $\epsilon_0$ by 
using  an attractive   or repulsive square well potential. The dependence of the signal-to-noise ratio $\Delta{\cal I}/{\cal I}_0$ and the 
square root of the variance of $\Delta{\cal I} \equiv {\cal J}$ respectively on the difference in binding energy between the ON and OFF states  
$\Delta \epsilon$ is displayed in Fig.\ref{fig:basalfretvar}. For low values of $\cal L$, the
effects on $\Delta{\cal I}/{\cal I}_0$ are not appreciable, however for larger values it is evident that varying -$\epsilon_0$ from -1 to 1 is 
reduced by more than half. $\cal J$ is high  where  $\Delta{\cal I}/{\cal I}_0$ is low and vica-versa, which is what one would expect intuitively.
$\cal J$  is sensitive to variations in low values of $\Delta \epsilon$, in particular when the basal interaction is repulsive. For moderate to 
high values of $\Delta \epsilon$ $\cal J$ is significantly lower  when the basal interaction is repulsive,  but only marginally in comparison with 
the neutral case of no interaction. 

\section{Conclusion and Outlook}
In this work general design features of  unimolecular FRET sensors were explored using simple coarse grained models and Monte Carlo Simulation. 
The starting point was the successful modeling of such sensors where highly optimized flexible linkers of \cite{komatsu2011} are used to connect 
the ligand binding and sensor domains. The flexible linker proteins  were then replaced by hinge like proteins where each arm is rod like, for 
example has the secondary structure of an alpha-helix. This  allowed four general design questions to be considered. First, can a simple 
mechanistic model of the \cite{komatsu2011} sensor capture the salient features observed in 
experiment, which we responded to in the affirmative. Second, is there an advantage in replacing the flexible linker of \cite{komatsu2011} with a 
hinge peptide? Here we were able to show that in general hinge peptides give far better results, except where the binding energy of the ligand 
binding and sensor domains is extremely low, in which case the performance is similar.
Third, to enhance precision of measurement, is it in  principle better to increase of decrease the the F\"oster radius of fluorescent proteins?
For flexible linker and hinge linker bases sensors, we saw that reducing the F\"oster radius can greatly enhance performance. 
Fourth, is precision enhanced or reduced if the binding energy of the ligand and sensor domains is attractive or repulsive in the absence of the 
target ligand? This turns out to depend on whether the binding energy between ligand binding and sensor domains is low of very high, and whether 
one focuses on the Signal to Noise ratio, or $\cal J$ (which is directly related to the Z' factor). For very high binding energies, $\cal J$ is 
not very sensitive, whereas the SNR is far more sensitive. As $\cal J$ is a better indicator of the quality of a sensor (lower values being 
better), for sensors having high binding energies, this is not a design concern to be overly concerned about. 

An alternative approach to enhance sensor performance is to choose hinge linkers which are  biased to be open
in the absence of the ligand through suitable choices of charged residues, so as to reducing false positive measurements. Results on that approach 
will be reported elsewhere.

\section*{Acknowledgment}
The work of DM is supported by the European Union under grant Number 676531 corresponding to the H2020 E-CAM Centre of Excellence.

\appendix

\section{Modeling the flexible linker}
\label{SEC:A1}
\begin{figure}[!htb]
\centering
\includegraphics[height =4 cm, angle = 0, width=6.0cm]{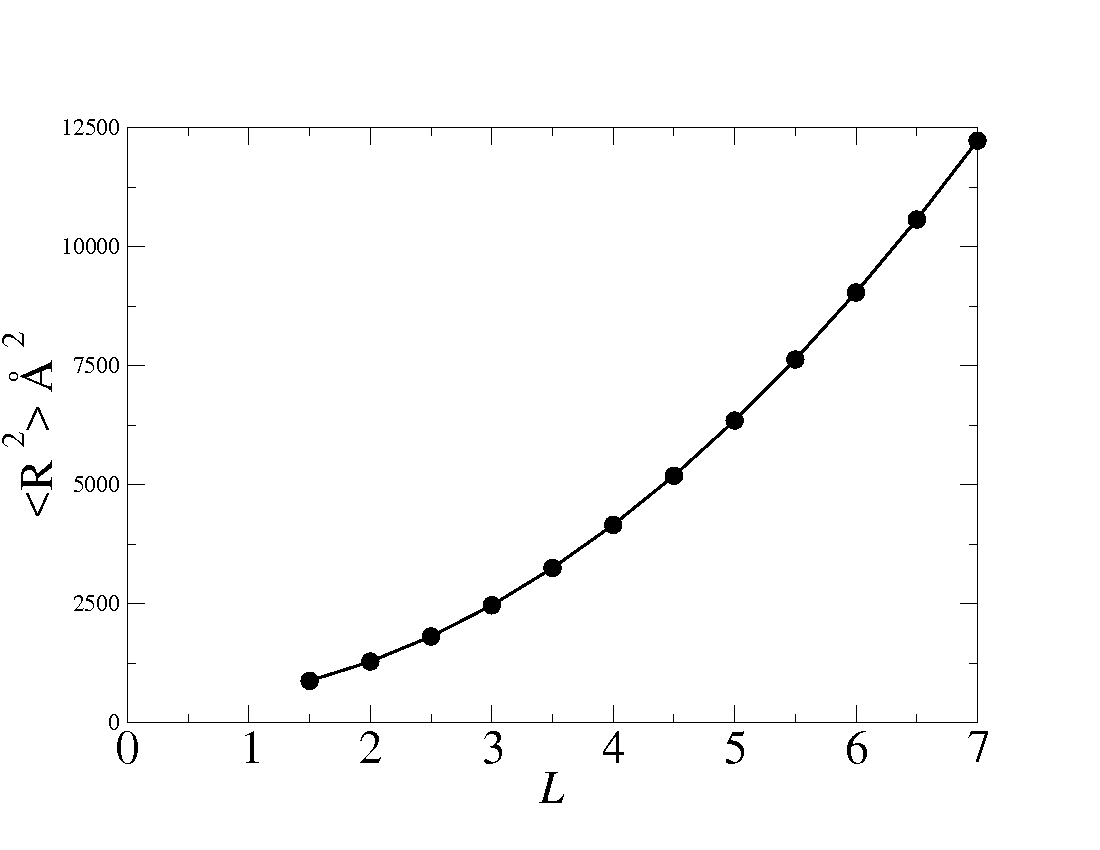} a
\includegraphics[height =4 cm, angle = 0, width=6.0cm]{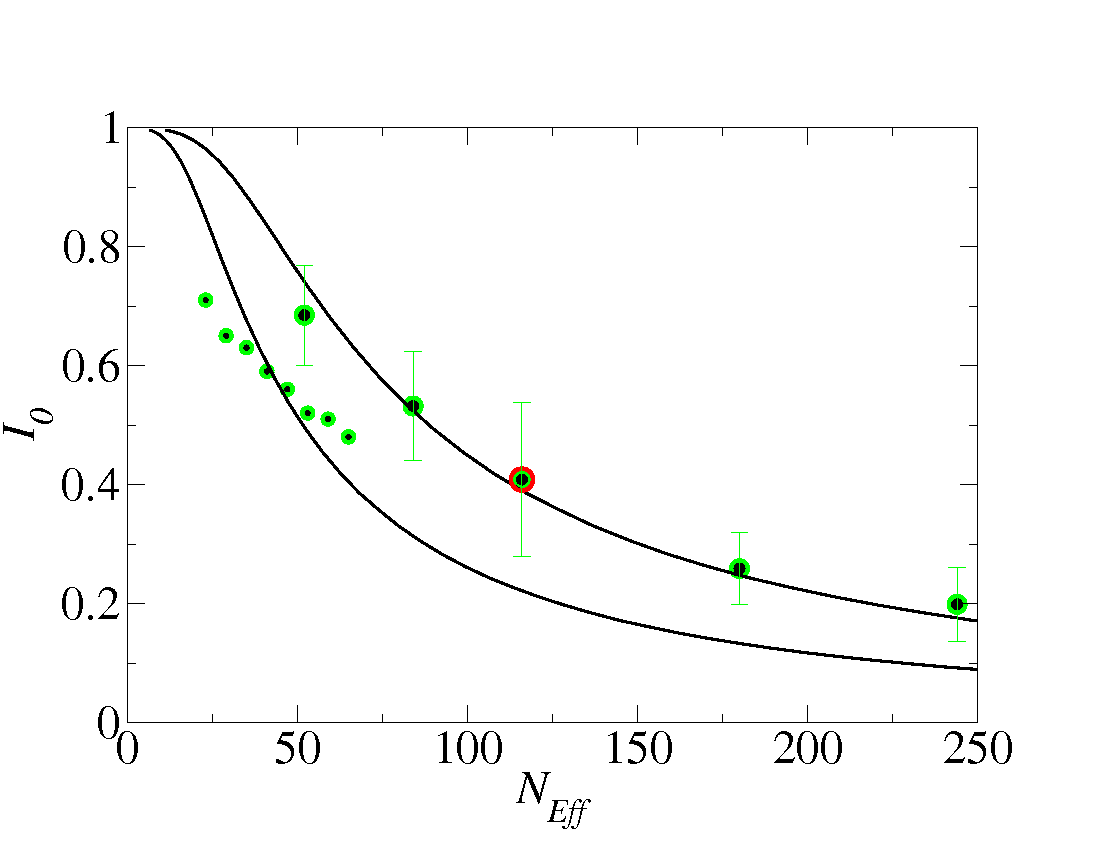} b
\caption{(a) Mean square displacement $<R^2>$  as a function of  $\cal L$. 
(b).  FRET sensitized donor-to-acceptor intensity ratio ${\cal I}_0$ as
a function of number of linker residues. Results presented are for data from experiments in the OFF or basal 
state from (1) Komatsu {\em et al}. (large filled green circles) and (2) Evers et al. (small filled green circles)  
superimposed on the theoretical predictions.}
\label{fig:experimentbasal}
\end{figure}

To compare resonance energy transfer (RET)  efficiency predictions of the simple flexible linker model where the parameter  $\cal L$
used in our model of the flexible linker, with experiment where the linker length is proportional to the  total
number of residues/beads, it is 
 necessary to relate $\cal N$ to $\cal L$. This is done using   the basal case (OFF state), by simply plotting the  mean square end to end
distance $<R^2>$ for the model and the experimental system respectively, where $r$  measured in units of $\sigma$ is converted $R$ measured in
units of $\AA$. An excellent fit to the data is given by
\[
<R^2> = \big( 274.89+251.61 {\cal L}^2 \big) \AA^2,
\]
as is evident in the fig. \ref{fig:experimentbasal} (a).  
For the experimental system, if the linker is sufficiently flexible, the end to end
displacement can be approximated as a Gaussian random walk,
\[
<R^2> = D_0 + C_{\infty} {\cal N} b_0^2,
\]
where $D_0$ is the square of the diameter of the macro-particle (the macro particles are assumed to have an effective diameter of  $ 24 \AA$),
$C_\infty  =3$ is the characteristic ratio and , $b_0 = 2.8 \AA$. Using,
\[N 
\sim \frac{274.89+251.61 {\cal L}^2 - D_0}{ C_{\infty} b_0^2}
\] 
to transform the dependence of the RET efficiency on $\cal L$ of the
model to the equivalent dependence on $\cal N$, we find excellent agreement with the corresponding experimental results of Komatsu {\em et al}. 
The only 
slight differences occurring when the linker is short, as discussed by  Evers et al\cite{Merkx2006}, and as a consequence the experimental results
depart from being an ideal Gaussian chain, where we have  used a     characteristic ratio $ C_\infty$ = 5. 
To compare the RET intensity of the ON state between the model and experiment, we use the scaling relation
of the basal case. It is worth pointing out that we have obtained
agreement also with more detailed models of the flexible linker. 

\begin{figure}[!htb]
\centering
\includegraphics[height =6 cm, angle = 0, width=9.0cm]{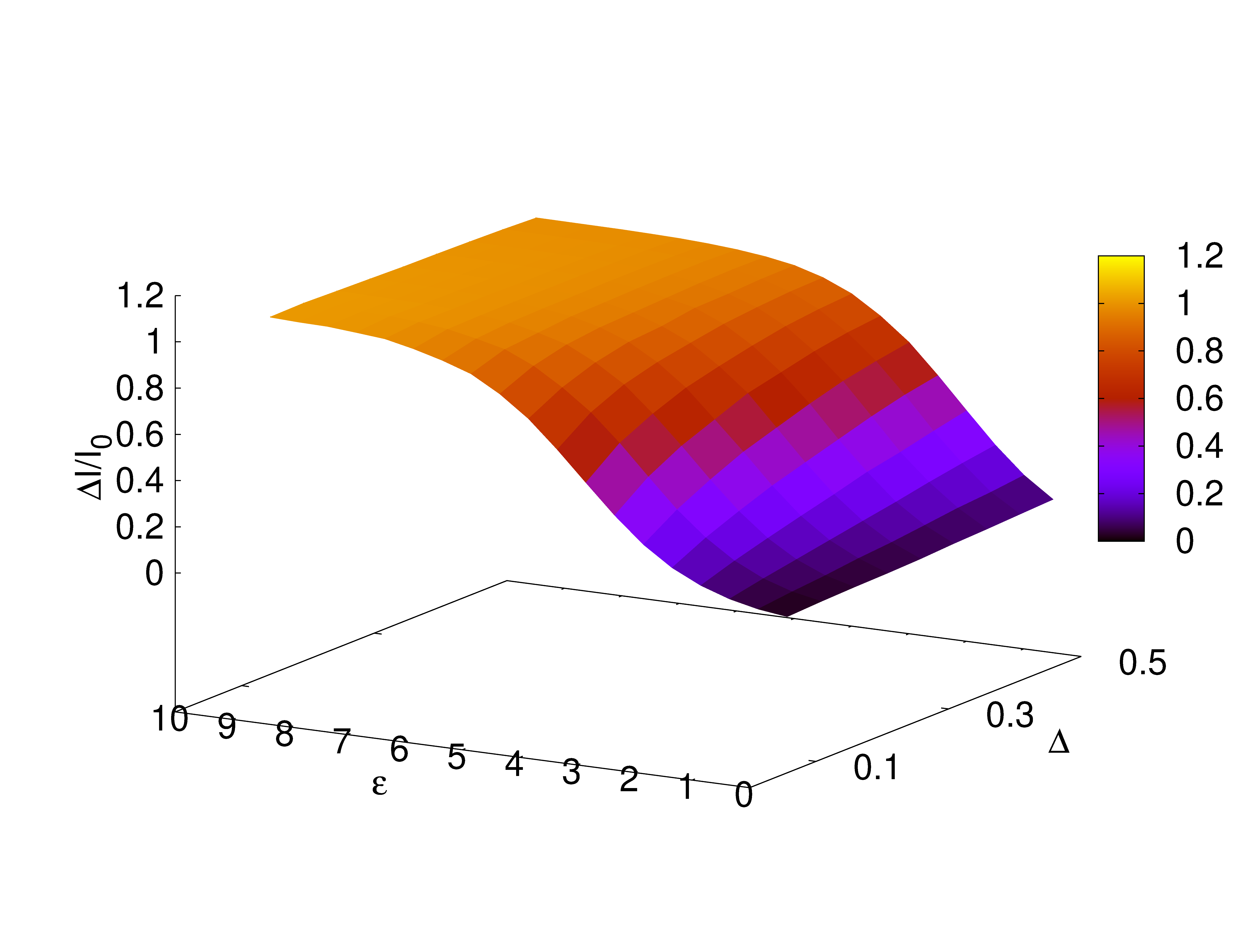} 
\caption{Dependence of the signal to noise ratio on the depth $\epsilon$ and width $\Delta$ of the attractive interaction (square well).
While $\Delta{\cal I}/{\cal I}_0$ has a strong dependence on $\epsilon$, there is little dependence on $\Delta$.}
\label{fig:flexible-extract-epsilon}
\end{figure}

One of the striking features of panel (a) of Fig.\ref{fig:flexible} in the main part of this paper was how easy it is to read off the
binding energy corresponding to the experiment of \cite{komatsu2011} by comparing their data with simulation. But in the ON state, 
the width $\Delta$ of the binding region as well as the depth $\epsilon$ can in principle influence the RET efficiency. To
investigate this issue we have simply varied both parameters in the theoretical model, the results of which are given  in Fig.
\ref{fig:flexible-extract-epsilon}.  We see, as one  might expect on theoretical grounds, that the signal to noise ratio has very little
dependence on $\Delta$,
which  also is therefore the case for the RET efficiency (as the basal rate can have no such dependence). 

\section{RET efficiency Observable and Sampling Procedure}
\label{SEC:A2}
The distance dependence of the FRET efficiency is approximated by the expression,
\[
{\cal I}(r) = {1 \over 1 + (r/R_{0})^6}
\]
with the F\"oster Radius $R_{0} \sim 5-7$ nm giving the distance at which the energy transfer efficiency is $50\%$ and 
$r$ is the distance between the spherical macro-particles.  The expectation value of the RET efficiency $\overline{\cal I}$ can be calculated
as an equilibrium average corresponding to the ON and OFF states respectively
\[
 \overline{\cal I} = \frac{\int dr {\cal I}(r)\exp(-\beta V(r))}{\int dr \exp(-\beta V(r))}
\]
where the multidimensional nature of $r$ is implicit.
$R_0$  depends on various quantities including  fluorescence quantum yield of the donor in the absence of the acceptor, 
the refractive index of the medium, and the dipole orientation factor $<\kappa^2>$. The orientation dependence is given 
as ${R_0}^6 \propto  <\kappa^2>$, where  $<\kappa^2>$ depends on the transition dipoles of the donor and acceptor 
fluorophores $\vec{D}$ and $\vec{A}$, and their mutual displacement $\vec{R}_{21} =\vec{R}_2 - \vec{R}_1$,
\begin{equation}
\kappa =  \vec{A} \cdot \vec{D} - \frac{3}{|\vec{R}_{21}|^2} \big( (\vec{D} \cdot \vec{R}_{21})( \vec{A} \cdot 
\vec{R}_{21}) \big)
\end{equation}
\label{eq:kappa}
If the two fluorophores rotate freely one can show that $\overline{\kappa}^2 = \frac{2}{3}$. This can be used to 
re-express
the efficiency in terms of the rotationally averaged  F\"oster radius $\overline{R_{0}}$  convenient for computation
\begin{equation}
{\cal I}(r) = {1 \over 1 + (r/\overline{R_{0}})^6 \frac {2 }{3} \frac{1}{<\kappa^2> }}
\end{equation}
\label{eq:r0-easier}
where the dependence of $\kappa$ on the transition dipoles and the mutual displacement of the fluorophores is implicit.

\clearpage





\bibliography{references_thesis_sj.bib}{}
\bibliographystyle{/media/3c0ee629-3374-4804-9313-1e7ca23c54a9/dmack/LATEX/PDFSLIDE/elsarticle/elsarticle-harv} 



\end{document}